\documentclass[11pt]{article}
\usepackage{moriond,epsfig}

\def\be{\begin{equation}}
\def\ee{\end{equation}}
\def\bea{\begin{eqnarray}}
\def\eea{\end{eqnarray}}

\begin{document}
\vspace*{4cm}
\title{GROWTH OF SPIRALS: SECULAR OR DRIVEN BY MERGERS ?}

\author{ F. HAMMER }

\address{GEPI, Observatoire de Paris-Meudon, 92195 Meudon, France}

\maketitle\abstracts{The  physical phenomena contributing to the galaxy growth can be tested all the way to z= 1. Galaxy mass, extinction, star formation and gas metal abundance can be measured in a robust way, as well as the distribution of the galaxy morphologies. I discuss here the observational methods and their accuracy. Physical quantities can be evaluated with uncertainties much lower than 0.3 dex, if they are based on 2 sets of independent measurements. For example, at a given IMF, the star formation rate is well estimated by combining flux measurements of the extinction corrected Balmer line and of the mid-IR continuum.\newline 
Spiral mass growth had occurred from gas accretion and from merging. Gas accretion can explain at most half of the spiral mass growth: at moderate redshift, the numerous population of compact, merger and irregular galaxies requires another origin. A spiral rebuilding scenario is able to reproduce all the evolutionary trends observed since z$\sim$ 1, and could be at the origin of the present-day, numerous population of early type spirals.}

\section{Introduction}

\subsection{Some general considerations on methodology }

The present-day generation of instruments allows to study the intermediate redshift galaxies with almost similar details than what has been done on nearby galaxies in the last 30 years. Investigations of the galaxy mass, dynamics, abundances, stellar population, star formation efficiency can be done. Methodology is crucial and we can learn a lot from previous works made in the local universe. Astronomers are following two main approaches in observing directly the galaxy evolution, e.g. the galaxies at high redshifts.\newline

On one side, galaxy surveys are gathering large or very large sample of galaxies. The aims are to minimise statistical uncertainty (for 10000 galaxies, $\sqrt{N}$/N is 1\%) and often to overcome the important problem of cosmic variance. Large samples can be easily gathered on the basis of multiple-band photometry, which is assumed to provide the redshift from analyses of the spectral energy distribution. However this implies very crude assumptions for deriving physical parameters, and it can lead to an uncomfortable feeling that the sophisticated instruments used to study the local galaxies are useless for those at large distances. For example, several studies, based on photometric redshifts, are characterising absolute luminosities, stellar masses or star formation rates from the single photometry. In extreme cases, by modelling spectral energy distribution, one might believe to be able to derive the extinction properties and their stellar population mixing. While interesting, these methods are likely subjected to large errors: what could be the impact of a few low redshift objects erroneously assumed at high redshift ? Corresponding derivations of luminosity, stellar mass or star formation densities may be subjected to uncontrolled biases. Large redshift surveys can overcome this problem and provide better estimates of physical quantities. However while they can be very efficient in gathering 10000s of objects, they have difficulties to estimate most physical parameters. For example it is practically impossible, to determine the dynamics of complex distant galaxies with slit spectroscopy (see e.g. Flores et al, 2005), or to derive proper extinction or gas abundances with spectral resolution well below R=1000 (see Liang et al, 2004a). \newline

On the other side, some studies, based on small samples of galaxies, are adopting observational techniques similar to those used for nearby galaxies. They are certainly facing the small statistics issue (for 100 galaxies, $\sqrt{N}$/N is 10\%). Because of this, they have to be very efficient in controlling the selection effects related to the target selection. Having saying this, they are quite unique to probe, at the same time, dynamics, stellar populations, gas abundances, extinction and star formation rate. Why this is so important ? Because z$\sim$1 is already probing the universe at less than half its present age, changes in galaxy properties are expected. Indeed, we know that the morphological appearances of distant galaxies are far more complex than those illustrated in the Hubble sequence. To study their complexity is at least equally demanding of precise instrumentation than what has been used for local galaxies. 

\subsection{The galaxies which are responsible of the z$<$ 1 star formation density decline }

The origin of the star formation density decline since the last 8 Gyrs is still a
matter of debate. Studies based on spectral energy distributions (SEDs) of galaxies 
predict that 30\% to 50\% of the mass locked in stars in present day galaxies
actually condensed into stars at $z$$<$1 (Dickinson et al. 2003; Pozzetti et al.
2003; Drory et al. 2004). This result is supported by the integrations of
the star formation density especially when it accounts for infrared light (Flores et
al. 1999). Indeed the rapid density evolution of luminous infrared galaxies (LIRGs, forming more than 20 $M_{\odot}$$yr^{-1}$) suffices in itself to report for the formation of $\sim$40\% of the total stellar
mass found in intermediate mass galaxies (from 3 $10^{10}$ to 3 $10^{11}$
$M_{\odot}$, Hammer et al. 2005). As opposed to the idea of galaxy ``downsizing''
(Cowie et al. 1996) -- strong evolution only in the faint blue dwarf population  --
there is a growing consensus that most of the decline of the star formation density 
is indeed related to intermediate-mass galaxies (Hammer et al. 2005; Bell et al.
2005). \newline

Here I would like to focus the reader's attention on how unclear are the words "massive" or "dwarves". I propose here that massive galaxies have $M_{star}$ $>$ 3 $10^{11}$ $M_{\odot}$ (i.e. bigger than the Milky Way), and that dwarves are defined by $M_{star}$ $<$ 3 $10^{10}$ $M_{\odot}$. This is somewhat arbitrary (as any definition), although with these criteria, E/S0 dominate the massive regime, spirals dominate the intermediate mass regime, and irregulars become a significant population in the dwarf regime (see Figure 1). \newline

It is more and more convincing that massive E/S0s and dwarves could not be responsible of the star formation density decline, because the first were mostly in place at z=1, and the second contribute marginally to the global stellar mass or metal content. In the following I will focus on the population of intermediate mass galaxies, in which at least 2/3 of the present-day stellar mass is concentrated (Brinchman and Ellis, 2000; Heavens et al, 2004). Today, according to the morphological classified luminosity function (see Figure 1 and Nakamura et al, 2004), 53\% are early type spirals (earlier than Sbc), 27\% are E/S0,  17\% are late type spirals, and only few (3\%) are irregulars.\newline

This paper focuses on intermediate redshift galaxies for which I (1) review the most robust estimates of physical properties at redshifts up to z= 1; (2) discuss the existing scenarios which are proposed as global pictures of the evolution since the last 8 Gyrs and (3) propose some further tests which can be done in the near future.

\begin{figure}
\center
\psfig{figure=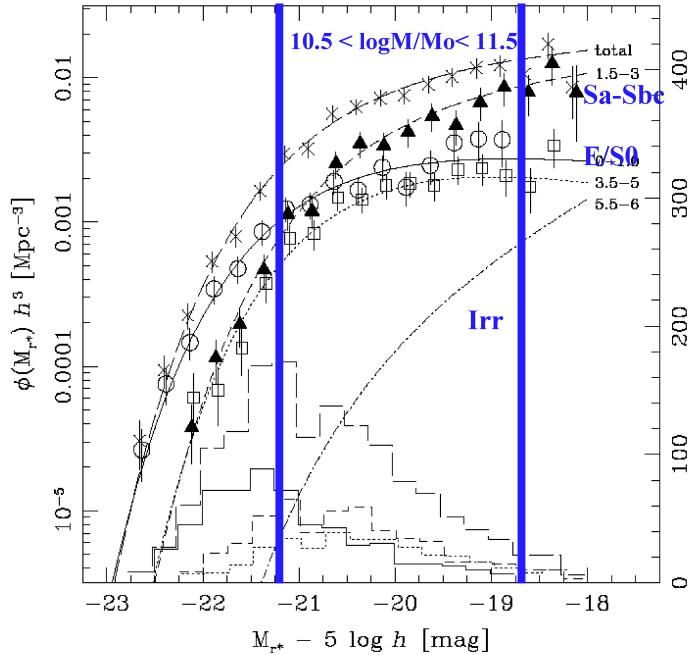,height=3.5in}
\caption{Adaptation of Figure 3 of Nakamura et al (2004) which present the morphologically determined luminosity functions (MDLFs) from the Sloan. The two vertical lines distinguish the area of intermediate mass galaxies. For this purpose I have assumed a ratio M/$L_{r*}$=5, by averaging  results from Glazebrook et al (2003) fits of the spectral energy distribution of SDSS galaxies.  
\label{fig:SDSS}}
\end{figure}

\section{Suggestions for a better estimation of the physical parameters at z=0.5-1}

Conversely to the local galaxies, most distant galaxies are barely resolved, and their physical properties can be only derived for the whole system. It might cause severe biases when comparing to local galaxies, because often those are observed through limited aperture systems (fibers or slits), which sample only a limited portion of the galaxy. More fundamentally we have to wonder what is the meaning of such "global" measurements of distant galaxy properties. As an example, abundances derived from $R_{23}$=$[O\,II]$$\lambda$3727/$H\beta$ + $[O\,III]$$\lambda$4959,5007/$H\beta$ can either sample all the ionised gas spreading over all the disk, or only a single HII giant region in which most of the star formation activity is concentrated. 

Nevertheless, all these measurements can be compared at different epochs and trace back the galaxy evolution. Below, I argue that at least 2 independent measurements (see Table 1) are necessary to evaluate, with a reasonable accuracy, the physical parameters of faint and distant galaxies.

\subsection{A major goal: galaxy mass estimates}

Of course, the most accurate method is dynamics. It is not easily applicable for distant galaxies because it is very demanding of photons (3D spectroscopy) or very uncertain if slits are used (how a slit has to be oriented for estimating the mass/dynamics of an irregular ?). Fortunately, the near IR luminosity of galaxies seems to correlate well with their stellar masses. This is intriguing because most of the near IR light is not coming from the main sequence stars which make most of the galaxy mass. It leads many authors to suspect that large uncertainties are related to these estimates, and wonder if these can apply equally to starbursts and early type evolved galaxies. Bell et al (2003) has ingeniously circumvented the difficulty, by applying an empirical correction depending on the B-V color of the galaxies (the bluer galaxies have lower stellar masses at a given $L_K$).  Nevertheless the uncertainty related to stellar mass estimates could be as high as 0.3-0.5 dex, without considering effects related to the IMF choice. Comparison with dynamical masses is required to examine how stellar mass estimates of starbursts compare to those of quiescent galaxies.

\subsection{Extinction, star formation rate and oxygen abundance}

Star formation rates (SFRs) are usually believed to be very uncertain. First factor of uncertainty is the supposed IMF: most (all ?) tracers of SFRs are linked with massive (IR) or very massive stars ($H\alpha$).  Kennicutt (1998) has provided us an useful tool, including formula deriving SFRs from various indicators, all assuming the same IMF. It is also relevant when comparing SFR with stellar mass to consider a common IMF for similar reasons of consistency. For practical reasons, UV or  $[O\,II]$$\lambda$3727 fluxes have been often used to characterise the  SFR, since these emission are redshifted to the visible window at moderate or high redshifts. Unfortunately, this does not work, especially at large distances, and let the SFR underestimated by very large factors (see Figure 2)!. Indeed  starbursts and LIRGs are so numerous at z$>$ 0.4, that the only viable tracers are those which account for the light reprocessed by dust (IR), or those properly corrected for extinction effect ($H\alpha$  after accounting for extinction using the $H\alpha$/$H\beta$ ratio). For a given IMF, mid-IR and extinction corrected $H\alpha$ fluxes can trace the SFR within an accuracy much better than 0.3 dex in logarithmic scale (Flores et al, 2004).\newline

\begin{figure}
\center
\psfig{figure=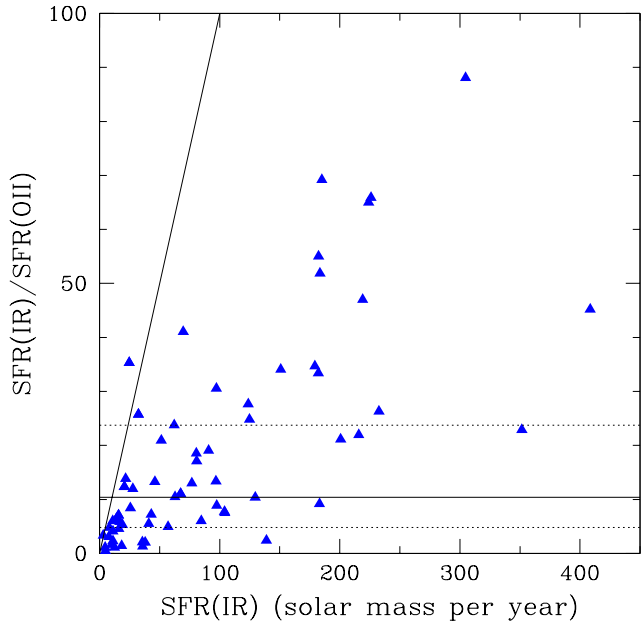,height=3.5in}
\caption{Ratio of SFR(IR)/SFR(OII) versus SFR(IR) for a sample of 67 CFRS galaxies 0.4$<$z$<$1.1, selected from Hammer et al (2005).  SFR have been calculated using the formalism of Kennicutt (1998), and same IMF for both estimates. All but one object show SFR(IR)$>$SFR(OII). The horizontal solid line show the median, and the dotted lines the upper and lower quartiles, respectively.
\label{fig:SFR_ratio}}
\end{figure}

\begin{table}[t]
\caption{Combination of measurements required for a robust determination of physical quantities of distant galaxies\label{tab:exp}}
\vspace{0.4cm}
\begin{center}
\begin{tabular}{|c|c|c|l|}
\hline
& & & \\
quantity &
1st measurement &
2nd measurement & notes
\\ \hline
galaxy mass & dynamics (3D spectrosc.) & stellar mass & error 0.3-0.5 dex on\\
& R $\ge$ 3000 & K band + $\ge$ 2 vis. bands &  stellar mass\\ \hline
gas extinction & Balmer line ratio & IR/visible ratio & error $<$ 0.3 dex \\ 
& corrected for absorption & aperture corrected &  \\ 
& R $\ge$ 1000& & \\ \hline
SFR & mid-IR & $H\alpha$ or $H\beta$  &  error $<<$ 0.3 dex\\ 
& & extinction corrected &  \\ \hline
gas abundance & extinction corr. $R_{23}$ & $ [N\,II]$/$H\alpha$ &  error $\sim$ 0.1 dex\\ 
& R $\ge$ 1000 & &  \\ \hline
morphology & "eyeball" classification & automatic softwares &  $\ge$ 2 indep. analyses \\ 
& & &  k correction indep. \\ \hline

\end{tabular}
\end{center}
\end{table}

The strong evolution of the LIRG number density evidences the need to proper account for extinction when estimating SFR or metal abundance using $R_{23}$.  $R_{23}$ is sensitive to the extinction through the ratio $[O\,II]$$\lambda$3727/$H\beta$. A difficulty is coming from the extinction calculation itself. At moderate redshifts, galaxies host significant populations of intermediate age stars responsible of strong Balmer absorption lines (see e.g. Hammer et al, 1997), which can easily affect the measurements of Balmer emission lines. One needs a sufficient spectral resolution (R$\ge$ 1000) and S/N ($>$ 10) to properly correct the underlying absorption, especially for measurements of $H\gamma$ and $H\beta$ lines. Another difficulty is due to the fact that, at z $>$0.4 the $H\alpha$ line is redshifted to the near-IR, and the $H\alpha$/$H\beta$ ratio has to be estimated using 2 different instruments. A way to circumvent the difficulty (and the cost of near-IR spectroscopy), is to use the ratio $H\gamma$/$H\beta$, although the $H\gamma$ line is often faint. Until now, very few studies are following these prescriptions, and caution has to be made in interpreting their results. The most exhaustive study has been made by Liang et al (2004b), who have systematically compared their extinction from $H\gamma$/$H\beta$ to that from the ratio of IR to $H\beta$ emission. A "nec le plus ultra" is to compare such robust results on $R_{23}$ to those from the ratio $[N\,II]$/$H\alpha$. This ratio is measuring nitrogen rather than oxygen and can be affected by the production of primary nitrogen (see Pettini and Pagel, 2004). However $R_{23}$ and  $[N\,II]$/$H\alpha$ seem to correlate well, at least for abundances above 1/4 of the solar values. $[N\,II]$/$H\alpha$ measurements can also help to resolve the degeneracy of the O/H-$R_{23}$ relationship.

\subsection{Morphological parameters}

Many attempts have been done to investigate the morphological evolution of galaxies. Although it has been established a long time ago that galaxy morphology indeed evolves, the variety of the adopted methods let the situation rather unclear. It is widely acknowledged that comparison of morphologies at different redshifts should be done at the same rest-frame: a spiral, if seen at UV wavelengths, could easily take the appearance of an irregular! Besides this, some classifications are based on eye analysis, other are based on automatic classifiers. The situation is even more confused because the various authors do not follow a common scheme for the galaxy classification. I propose to define the different classes as simple as possible (following  Zheng et al, 2004, 2005), and to classify galaxies using only 4 categories: E/S0, spirals, irregulars and objects too compact to be securely classified in the 3 first categories. Because the fraction of compact galaxies is rather high at intermediate z, attempt to force their classification following the Hubble diagram can lead to uncontrolled biases. To minimise the uncertainties, it is crucial to combine here independent examinations ("eyeball") by experimented astronomers, with results from automatic softwares. Colour maps also help to minimise the effect of k-correction.

\section{The prominent features of the galaxy evolution}

\subsection{Evolution of luminosity densities}

The common decline of the UV and IR luminosity with the epoch is widely accepted, although its magnitude is still a matter of debate. From z=1 to z=0, the UV luminosity decreases by a factor 5 to 10 (Lilly et al, 1996; Cowie et al, 1996), while the IR luminosity decreases even more rapidly (see Elbaz and Cesarsky, 2003).  Such a decline is followed by a strong diminution of the fraction of emission line galaxies (defined as those with $W_{0}(OII)$ $\ge$ 15 A), from $>$ 70\% at z=1 to 17\% at z=0, e.g. Hammer et al, 1997). These changes are related to a general decline of star formation in most galaxies during this period (Madau et al, 1996) with a marginal contribution of AGN.If accounting for the IR luminosity density evolution, the integration of the derived SFR matches well the reported evolution of the stellar mass density.

Hammer et al (2005) has shown that most of the stellar mass formation since z=1 has occurred in LIRGs, which evolve the most rapidly: 15\% of intermediate mass galaxies were LIRGs at z$\sim$ 0.7 which compares to only 0.5\% today. Because LIRGs show high specific SFR (SFR/$M_{\star}$ $\sim$ $(0.8Gyr)^{-1}$), Hammer et al propose that they correspond to episodes of enhanced star formation in most intermediate mass galaxies. The star formation in LIRGs is sufficient in itself to produce 38\% of the present-day stellar mass of intermediate mass galaxies. It then accounts for most of the reported stellar mass formation since z=1 (see also Le Flo'ch et al, 2005).

\subsection{Some clues on the evolution of the gas available to star formation}

Estimates of the gas metallicities in distant galaxies are becoming more and more accurate. Former estimates by Kobulnicky and Kewley (2004) and Lilly et al (2003) were limited by the absence of flux measurements and extinction estimates, respectively (see Liang et al 2005 for more information). More recently, Liang et al (2005) and Savaglio et al (2005) have established the stellar mass-metallicity diagram for a set of intermediate mass galaxies and dwarves, respectively. In both studies, a significant evolution is found, z $\sim$ 0.7 galaxies showing on average, gaseous abundances 2 times smaller than that of local counterparts at a given stellar mass (see Figure 3). Such an effect is also found by Maier et al (2005), although with a smaller amplitude. 
Notice however that in Maier et al, NII/$H\alpha$ provide systematically lower abundance values than that from $R_{23}$, which could reflect either a real evolution of the nitrogen production or a bias.\newline

Using a proper account of the extinction effects (see section 2.2), Liang et al (2005) find that, within all the range of  $10^{10}$ $M_{\odot}$ to  3 $10^{11}$ $M_{\odot}$, galaxies show similar metal deficiencies. If true, this indicates a gas fraction available in z=0.65 galaxies more than twice  that in local galaxies (assumed to be 10-20\%).  A closed box model predicts a gas fraction converted into stars of 20-25\% since z$\sim$ 0.65, in excellent agreement with findings based on the evolution of the stellar mass density (see e.g. Drory et al, 2004). Of course, a closed box model is a very crude
approximation of the evolution of intermediate mass galaxies, which can be
affected by gas outflows, inflows and minor or major mergers. However, as a whole,
intermediate mass galaxy population can be assumed to characterize most of the gas
to mass ratio of the universe, and applying a closed box  model to this population
allows comparison to other integrated quantities such as stellar mass or star
formation densities. 

The absence of slope change of the $M_{star}$-Z
relation found by Liang et al (2005), somewhat contradicts the "downsizing" scenario. This needs to be confirmed by measurements on larger data sets, following the method outlined in section 2.2.

\begin{figure}
\center
\psfig{figure=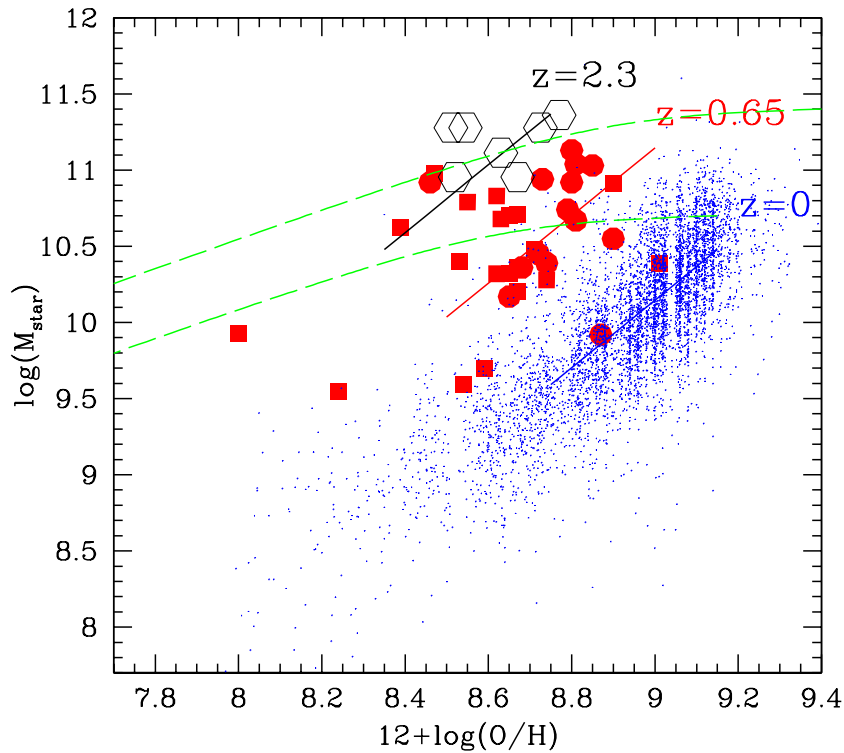,height=4.5in}
\caption{Reproduced from Liang, Hammer and Flores (2005). Stellar mass-metallicity relation for the SDSS galaxies (z=0, Tremonti et
al, 2004), the intermediate-$z$  galaxies (z$\sim$0.65, red points from Liang et al, 2005) and the 7 galaxies (black open hexagon) at z=2.3 described by Shapley et al.
(2004).   By reporting the lines to fit the median of intermediate distant
galaxies and of z=2.3 galaxies, we find an evolution of -0.35 dex and -0.7 dex in
metal abundances, respectively. The green long-dashed lines show the result of a
closed box model and indicating the general evolutionary trend from z=2.3 and
z=0.65 to z=0, respectively (see text). 
\label{fig:M-Z}}
\end{figure}

\subsection{Morphological evolution}

All studies of galaxy morphologies have revealed a clear evolution which is mainly due to the emergence, at moderate redshifts, of peculiar, compact or irregular galaxies in the intermediate mass range (Brinchmann et al, 1998; Lilly et al, 1998; Abraham and van den Bergh, 2001). Table 1 presents the morphological classification established by Zheng et al (2004, 2005) of a sample of 111 intermediate z galaxies, including LIRGs. It has been done following the method described in section 2.3, and it is consistent with the above mentioned investigations of galaxy morphology. Notice that 64\% of LIRGs are made of peculiar, compact or irregular galaxies, which further supports the key role of these galaxies in the galaxy evolution. However this fraction is based on only 36 LIRGs: recently, Melbourne et al (2005) find a smaller value (50\%), after analysing 119 LIRGs in the GOODS-N field.

\begin{table}
\caption{From Hammer et al (2005): morphological classification statistics for intermediate mass galaxies; z $\sim$ 0.7 galaxies are compared to those of the SDSS (Nakamura et al, 2004).}
\begin{tabular}{lccc}
\hline
Type   & z$\sim$0.7 & z$\sim$0.7 & local \\ 
       & LIRGs &  galaxies & galaxies \\ \hline
E/S0        &   0\% & 23\% & 27\%\\
Spiral    &  36\% &  43\% & 70\%\\
LCG  &  25\% &  19\% & $<$ 2\%\\
Irregular  &  22\%  &  9\% & 3\%\\
Major merger &  17\% &  6\% & $<$ 2\%\\ \hline
\end{tabular}
\end{table}

\section{Towards a global picture of the galaxy evolution since z=1}

Because a significant fraction of present day stars have been formed in LIRGs since the last 8 Gyrs, the LIRG morphology is a key issue. Bell et al (2005a) are finding that more than half of LIRGs in the GOODS-S field are apparently undisturbed spirals, and conclude that most of the recent star formation was occurring in regular spiral galaxies. Bell et al (2005a) also find that 10\% of LIRGs are E/S0, which are not or marginally detected ($<$ 1\%), either by Zheng et al (2005) or by Melbourne et al (2005). Differences between these 3 studies might be due to cosmic variance: indeed the GOODS-S field show very peculiar properties (see Le Flo'ch et al, 2005) with two prominent peaks in the galaxy redshift distribution at z=0.67 and z=0.73, those being included in the Bell et al study. Given this and the statistical uncertainties related to all these studies, it is likely that the fraction of spirals which are LIRGs is 50$\pm$10\% and equals that of peculiar/compact and irregular galaxies. Dynamics from 3D spectroscopy is the only way to precise this value, i.e. which fraction of LIRGs are indeed rotating disks (Flores et al, 2005). Let us now examine the different propositions which relate the z=1 to z=0 intermediate mass galaxies. 

\subsection{A rebuilding disk scenario ?}

The high occurrence of LIRGs is easily understood only if they correspond to episodic peaks of star formation, during which galaxies are reddened through short IREs (infrared episodes). Recall that 15\% of intermediate mass galaxies at z$\sim$ 0.7 are LIRGs, a fraction which, according to Melbourne et al (2005), increases to 75\% of the luminous $M_B$$<$-21 galaxies. Given the high specific star formation rate of LIRGs, it is difficult to avoid the laternative of an episodic star formation: otherwise, LIRGs would be progenitors of massive ($M_{star}$ $>$ 3 $10^{11}$ $M_{\odot}$) galaxies, multiplying their masses by a factor close to 10, since the last 8 Gyrs (see Hammer et al, 2005)! \newline

An episodic star formation is highly suggestive of expectations from hierarchical building of galaxies, where peaks of star formation corresponds to galaxy merging. Hammer et al (2005) argue that major merging should play an important role to account for the high fraction of compact galaxies and major mergers (see Table 1). Present-day galaxies are mostly spirals (75\%), E/S0 show no evidence for a strong number evolution, and the rapid evolution of compact/major merger and irregular is plausibly linked with spiral progenitors. It is simply excluded that compact galaxies could be progenitors of ellipticals, given the relative number densities shown in Table 1. \newline

Hammer et al (2005) propose an evolutionary sequence of disk galaxy formation, in which a significant  fraction ($\sim$75\%) of spirals have experienced a major merger able to destroy the disk of the progenitor. The post merging phase would correspond to compact galaxies, for which most of the gas and stars have fallen to a common barycentre, and then, it is followed by the disk rebuilding. Because of the feedback, it is likely that the whole process can be observed during more than 1 or 2 Gyr, showing all the morphologies described in the 2nd column of Table 2. It is remarkable that this scenario suffices to account for the morphological changes, as well as for the dramatic evolution of UV and IR luminosity densities (including LIRGs).  \newline

Recent simulations (Robertson et al, 2005; Springel and Hernquist, 2005; Di Matteo et al, 2005) are challenging the conventional origin of disk galaxies, and show that many spirals could be indeed
formed after major merger events. It brings a significant support to the spiral rebuilding hypothesis: due to the feedback by supernova and AGN, a significant fraction of the gas could be expelled towards the galaxy outskirts, allowing a gas reservoir for the elaboration of a new disk (see Figure 4). It can also solve the so called angular momentum problem: Robertson et al (2005) demonstrate that if large feedback were occurring, they may lead to rejuvenated disks with high values of the angular momentum. \newline

\begin{figure}
\center
\psfig{figure=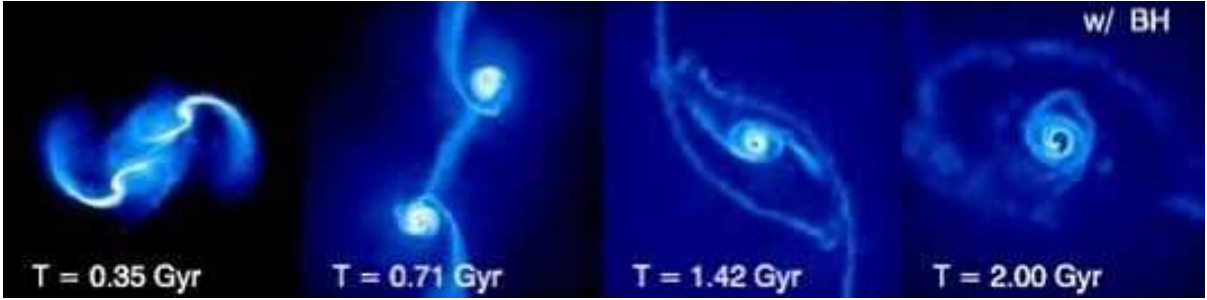,height=1.55in}
\caption{Reproduced from Robertson et al simulations (their Figure 1). Merging of two gas rich galaxies with equal masses with blachole and supernova feedback. It shows the gas surface" density over 140 x 140 kpc, demonstrating that the high-angular momentum merger results in a remnant disk. The effect of the feedback can be seen over the 2 Gyr time scale of the merger, as a diffuse, hot halo, and a reduced central density compared to simulations without feedback.
\label{fig:M-Z}}
\end{figure}

This scenario has to deal with the questions of the apparently unchanged number density of disks at z=1 (Lilly et al, 1998), the redshift increasing fraction of apparently normal spirals in LIRGs (Melbourne et al, 2005), and not the least, the fact that this scenario is not applicable to our own Galaxy. The first question should deserve more investigations (simulations ?), although it is unclear if all morphologically identified spirals at high z are indeed rotating disks (but see Flores et al, 2005). A significant fraction of major mergers would unavoidably lead to a decrease of the number of spirals with time, which can be somewhat compensated by the merging of more smaller units which then enter the intermediate mass range. \newline

The disk rebuilding scenario is also extremely dependent on the gas availability, which was likely much higher in the past than today (see section 3.2). Disk rebuilding is plausible only if there was enough gas available. Indeed, the gas being exhausted with time, it would unavoidably limit the star formation in the post merger phase at low redshift. No more gas being available to form a new disk, one can expect that no more forming disks with strong star formation could appear at low redshift, as it is reported by Melbourne et al (2005). \newline

A scenario for which our own Galaxy is not an archetype might appear unattractive.  In fact the rebuilding scenario is particularly well suited to explain the large fraction of early type, bulged spirals (75\% of present-day spirals are earlier than Sbc, according to Nakamura et al, 2003). If those bulges have been formed during the last 8 Gyrs, it let the Milky Way as one of the (late-type) spirals which have escaped a major encounter since that epoch. Let us consider the other nearby, intermediate mass galaxy for which robust conclusions on the past formation history can be derived , e.g. M31. There are several new evidences that the M31 past history was much more perturbed than that of the Milky Way. Among them, there are the presence of a significant population of 6 Gyr stars in the M31 halo (Brown et al, 2003; Rich, 2004) and the small age of the M31 thin disk (1-2 Gyr, see Beasley et al, 2004). Those can be easily accounted in the frame of the rebuilding spiral scenario if a major merger had occurred in M31, about 6 Gyr ago.

\subsection{A secular evolution mixed with gas infall ?}

It has been advocated for a long time that disks were experiencing secular evolution with an almost constant star formation (Kennicutt, 1992). Bournaud and Combes (2002) have shown that, during a cycle of bar formation and destruction due to the gas infall, spirals can double their masses during a Hubble time. It also allows to form bulges, although it is questionable if bulges of early type spirals can be formed this way (Debattista et al, 2004). A secular evolution associated with gas infall has been revivified by Dekel and Birboim (2005) to interpret the bi-modality of the properties of present-day galaxies. \newline

Observations of a large density of distant spirals at z$\sim$ 1 may bring a significant support to a secular evolution. However it has to be accommodated with an episodic origin of these gas accretions.  An important question is the origin of the gas, which can come from the filaments or from the infall of several dwarves during the Hubble time. Those alternatives let open the issue of the high angular momentum observed in present-day disks. 

Another support to a gas infall scenario is provided by some observations of the numerous infrared galaxies which, even sometimes with SFR larger than 100 $M_{\star}$$(yr)^{-1}$, show the regular appearance of large disks. More recently, Flores et al (2005) show that a few of them are normal rotating disks, which let open the possibility that large gas infall are replenishing the disk. A major limitation of this scenario is caused by the strong evolution of the galaxy morphologies, and it can explain at most half of the star formation during the last 8 Gyrs, a fraction corresponding to that of spirals in LIRGs.   

\section{Conclusion and further tests}

As shown above, many significant progresses have been recently done towards the understanding of the galaxy formation. Recall that less than 5 years ago, it was often believed that intermediate mass and massive galaxies were almost "dead" during the last 8 Gyrs. IR observations with ISO and Spitzer have revealed a much more lively picture, showing dramatic evolution and episodic star formation history. The major scenarii of galaxy evolution are getting more closer today, with a significant fraction of star formation related to both gas infall and galaxy merging. In fact the debate between the two alternatives resembles what we call in french "le d\'ebat sur la bouteille \`a moiti\'e vide ou \`a moiti\'e pleine". \newline

There were more gas available to galaxy formation in the past, as evidenced by the gas metal evolution and by the considerable evolution of LIRG number density. A fundamental question is to identify where the gas was lying before being transformed into present-day stars. The episodic star formation revealed by the high LIRG occurrence is difficult to reconcile with a smooth and low density gas distribution such as that lying in the filaments.\newline

Important progresses are needed to disentangle the impact of the different physical mechanisms in the formation of bulge and disk of the vast majority of galaxies, e.g. the spirals. One issue is the frequency of major mergers which is still a matter of debate, although the net majority of investigators (e.g. Le F\`evre et al, 2000; Conselice et al, 2003; although see Lin et al, 2005) found a declining merger rate, implying from 0.5 to 1.5 major merger per L* galaxy.  If true, this gives further credit to the spiral rebuilding scenario of Hammer et al (2005). The main difficulty to evaluate the major merger rate is caused by the fundamental uncertainty in estimating the relative mass ratio between the encounters (see however Bundy et al, 2004 and Rawat et al, 2005, in preparation). \newline

Another question is related to the little evolution seen in ellipticals which seems so different to what is observed in spirals. Bell et al (2005b) have shown that ellipticals can be also affected by major merging, at a similar rate than spirals. However because ellipticals are likely gas empty, these events can be described as dry mergers, i.e. without being associated to strong star formation events. It might seem that Bell et al (2005b), by supporting the dry merging scenario for elliptical is contradicting Bell et al (2005a) who is arguing against a merger scenario for the spiral mass assembly. This could be solved in a way or another, by investigating if, within a similar mass range, gas poor galaxies lie in much richer environments than gas rich galaxies.\newline

A major issue is to investigate further the dynamics of apparently normal spirals at high redshifts. Are those normal rotating disks in equilibrium ? New observations with the multi-integral field unit mode of FLAMES/GIRAFFE at the VLT will soon shed new light on this issue (see Flores et al, 2005).  Velocity fields of compact galaxies will also test if they correspond to a post-merger phase as it is assumed by the spiral rebuilding scenario (see Puech et al, 2005). A study of the evolution of bulge/disk ratio, independent of k-correction, is also necessary (Zheng et al, 2005, in preparation).\newline

Detailed comparisons with simulations are highly welcomed. Those simulations are now enough accurate to recover most of the physical properties of galaxies at any redshifts. Very few attempts have been done to simulate the observed evolution of the major merger rate. In their simulations, Murali et al (2002) find an almost constant merger rate all the way from z=1 to z=0, and concluded that most of the galaxy mass growth is due to gas accretion. It supports more the Lin et al (2005) results than those of most of the other investigators. It has to be clarified if this is a robust conclusion from cosmological simulations or not. A detailed simulation of the M31 past history would be extremely useful, as what has been often done for the Milky Way.

\section*{Acknowledgments}
I'd like to thanks the organizers, David Elbaz and Herv\'e Aussel for this wonderful conference and the very stimulating discussions which I have benefited there. I'd like to thank my collaborators for the work we have done together, and for the pleasure of working in such a stimulating atmosphere at the GEPI, Paris Observatory.

\end{document}